\documentclass{article}
\usepackage{spconf,amsmath,graphicx}

\usepackage[noend]{algpseudocode}
\usepackage{cite}
\usepackage[linesnumbered,ruled,vlined]{algorithm2e}
\usepackage{makecell}
\usepackage{tabularx}
\usepackage{float}
\usepackage{stfloats}
\usepackage{booktabs}
\usepackage{array}
\newcolumntype{C}[1]{>{\centering\let\newline\\\arraybackslash\hspace{0pt}}m{#1}}

\usepackage[colorlinks=true,citecolor=green]{hyperref}
\usepackage{cleveref}

\usepackage{caption}
\usepackage{subcaption}
\usepackage{adjustbox}
\usepackage{subfiles}
\usepackage{amssymb}
\usepackage{pifont}
\usepackage{multirow}
\usepackage{xcolor}

\title{FAT-HuBERT: Front-end Adaptive Training of Hidden-unit BERT for Distortion-Invariant Robust Speech Recognition}
\name{\begin{tabular}{c}Dongning Yang, Wei Wang, Yanmin Qian$^\dagger$\thanks{$^\dagger$ corresponding author}\end{tabular}}
\address{
MoE Key Lab of Artificial Intelligence, AI Institute \\ Department of Computer Science and Engineering Shanghai Jiao Tong University, Shanghai, China \\
\{ydn\_1007,wangwei.sjtu,yanminqian\}@sjtu.edu.cn}
\copyrightnotice{979-8-3503-0689-7/23/\$31.00~\copyright2023 IEEE}

\begin{document}
%\ninept
%
\maketitle
\begin{abstract}
Advancements in monaural speech enhancement~(SE) techniques have greatly improved the perceptual quality of speech. However, integrating these techniques into automatic speech recognition~(ASR) systems has not yielded the expected performance gains, primarily due to the introduction of distortions during the SE process. In this paper, we propose a novel approach called FAT-HuBERT, which leverages distortion-invariant self-supervised learning~(SSL) to enhance the robustness of ASR. To address the distortions introduced by the SE frontends, we introduce layer-wise fusion modules that incorporate features extracted from both observed noisy signals and enhanced signals. During training, the SE frontend is randomly selected from a pool of models. We evaluate the performance of FAT-HuBERT on simulated noisy speech generated from \textsc{LibriSpeech} as well as real-world noisy speech from the \textsc{CHiME-4} 1-channel dataset. The experimental results demonstrate a significant relative reduction in word error rate~(WER).
\end{abstract}
\begin{keywords}
self-supervised learning, robust speech recognition, Front-end Adaptive Training, HuBERT
\end{keywords}
%

%%%%%%%%%%%%%%%%%%%%%%%%%%%%%%%%%%%%%%%%%%%%%
%%%%%%%%%%%%%%% Introduction %%%%%%%%%%%%%%%%
%%%%%%%%%%%%%%%%%%%%%%%%%%%%%%%%%%%%%%%%%%%%%

\section{Introduction}
\label{sec:intro}
% introduction

% Speech enhancement, self supervised learning
% 

% paragraph 1
% two dominating approaches for robust speech recognition
% 1. robust speech recognition backend
% 2. speech enhancement frontend

Robust speech recognition constitutes a pivotal area of study within the field of automatic speech recognition~(ASR) due to its capacity to significantly augment system performance in real-world, noise-prone environments~\cite{radford2023robust, zhang2018deep, hu2022interactive, ravanelli2020multi}. 
The primary objective of robust speech recognition is to enable accurate and efficient recognition of speech in the presence of diverse noise types and varying intensities, which typically interfere with the accurate extraction of linguistic information.

% paragraph2
% introduce robust speech recognition methods
% 1. conventional methods
% 2. self supervised learning methods, wav2vec2.0, hubert, cpc, c-siam, data2vec?, etc

Research on robust speech recognition can be broadly divided into front-end and back-end techniques, based on the stage at which system noise-robustness is integrated. The recent growth of self-supervised learning (SSL) methods, emerging as promising ASR back-ends~\cite{cpc, hubert, wav2vec2, decoar, c-siam}, has led to a wealth of proposed strategies to combat the susceptibility of SSL models to background noise within ASR tasks.
WavLM~\cite{chen2022wavlm}, for example, adopts a masked speech denoising and prediction framework for pretraining speech representations. Wav2vec-Switch~\cite{wav2vec-switch} predicts the quantized representations of the original-noisy speech pairs fed to wav2vec2.0~\cite{wav2vec2} network. Furthermore, Wav2vec-C~\cite{wav2vec-C} and \cite{wav2vec-recons} incorporate a reconstruction loss into the wav2vec2.0 framework. HuBERT-AGG~\cite{wang2023hubert} employs a distinct approach by distilling layer-wise noise-invariant representations to bolster the robustness of HuBERT~\cite{hubert}.

% paragraph3
% introduce speech enhancement frontend methods, introduce distortion problem, approaches to alleviate this problem: multi-style training, joint training of frontend and backend

% Regarding frontend techniques for robust ASR, a speech enhancement~(SE) module is often integrated as a pre-processing frontend to suppress the noise from noisy speech, where the SE and the ASR module can be trained separately or jointly. However, it has been observed in many previous works~\cite{iwamoto2022bad, mporas2010speech, wang2019bridging, loizou2010reasons} that the enhanced speech from SE might not always yield good recognition accuracy for the downstream ASR task which might be attributed to the distortions in terms of intelligibility of enhanced speech not optimal for ASR. 

Front-end techniques for robust ASR often employ a speech enhancement (SE) module as a pre-processing front-end to reduce noise within the speech signal. The SE and ASR modules can be trained independently or jointly. However, as numerous prior studies have observed \cite{iwamoto2022bad, wang2019bridging, loizou2010reasons}, the enhanced speech output from SE does not consistently translate into optimal recognition accuracy for subsequent ASR tasks, an issue often attributed to sub-optimal intelligibility distortions within the enhanced speech.
To alleviate this problem, \cite{wang2016joint, liu2019jointly} proposed the joint training of the SE and ASR modules using ASR objectives, thereby restricting the loss of linguistic information induced by distortions during SE. Moreover, it has been demonstrated in \cite{iwamoto2022bad, hu2022interactive, fan2020gated} that the fusion of features derived from both observed noisy signals and enhanced signals can effectively compensate for each other, resulting in features that are not only noise-robust but also less susceptible to distortion.

% paragraph 4
% in this paper, we combine se frontend and ssl backend
% list existing works, limitations of existing works,
In this study, we introduce a novel training approach for building distortion-invariant SSL models that adapt to a multitude of diverse pretrained SE front-ends using the HuBERT framework. We start with a pre-trained HuBERT, which is refined for distortion robustness without extensive additional training steps. During pretraining, the model is exposed to both observed noisy signals and enhanced signals, where the SE front-end is randomly chosen from a collection of SE models. We employ a layer-wise fusion mechanism that combines features from noisy and enhanced signals, resulting in features with reduced noise and distortion. Additionally, we propose intra-utterance multi-style training that partially enhances each utterance, effectively lowering GPU memory overhead and minimizing training speed degradation caused by SE front-ends. To safeguard the initialization parameters against premature continual pretraining, we incorporate a residual connection for each fusion module.
Our work distinguishes itself from \cite{gao2021multi}, which utilized a data-driven approach to calculate contrastive loss among various acoustic conditions. In our case, features from noisy and enhanced signals are deeply fused via a parameterized module within the SSL model. Moreover, unlike \cite{zhu2022joint}, which jointly optimized the SSL and SE module for distortion robustness, we retain the SE models' original state during training. Instead, we leverage a collection of SE front-ends to train a front-end adaptive SSL model. In contrast to the study in \cite{wang2019bridging} that used time-frequency~(TF) domain front-ends to train an acoustic model, we investigate both time domain and TF-domain front-ends to train a distortion-invariant SSL model.

% paragraph 5
% contributions

Our contributions can be summarized as follows:
(1) We introduce a front-end adaptive training scheme integrated with HuBERT~(FAT-HuBERT) that effectively mitigates distortions introduced by SE front-ends.
(2) We propose intra-utterance multi-style training that lowering GPU memory overhead and reduce the time required for data processing during FAT-HuBERT pretraining.
(3) Experimental results demonstrate that our FAT-HuBERT framework significantly improves the robustness of learned representations against noise and distortions, leading to substantial word error rate~(WER) improvement on the \textsc{LibriSpeech} simulated noisy speech dataset and \textsc{CHiME-4} 1-channel real-world noisy speech.

\section{Front-end Adaptive Training of HuBert}
\label{sec:methodology}

\subsection{HuBERT}
\label{subsec:hubert}
We first revisit Hidden-Unit BERT (HuBERT), the underlying model for our methodology. HuBERT is a self-supervised approach for latent representation learning, which demonstrates superior performance and generalization across varied applications. Utilizing an offline clustering step like K-Means, HuBERT aligns target labels for BERT-like prediction loss~\cite{bert}, predicting cluster assignments from masked speech features. A speech utterance $X=\left[x_1, \ldots, x_T\right]$ with a clustering model $h$ yields acoustic units $h(X) = Z = [z_1,\ldots, z_T ]$ with $z_t \in [C]$ as a categorical variable. The prediction loss applied only to masked regions impels a combined acoustic-language model over continuous inputs. 

HuBERT's architecture includes a convolutional waveform encoder, a BERT encoder, a projection layer, and a code embedding layer. The model $f$ processes a masked embedding sequence $\tilde{\mathcal{X}}=r(\mathcal{X}, M)$, derived from the length-$T$ CNN encoder output $\mathcal{X}$, to predict distribution $p_f(\cdot \mid \tilde{\mathcal{X}}, t)$ across target indices at timestep $t$, given by:

$$
p_f(c \mid \tilde{\mathcal{X}}, t) = \frac{{\rm exp}({\rm sim}(\phi(\tilde{\mathcal{X}})_t\mathbf{W}, \mathbf{e}_c)/\tau)}{\sum_{c'=1}^{C}{\rm exp}({\rm sim}(\phi(\tilde{\mathcal{X}})_t\mathbf{W}, \mathbf{e}_{c'})/\tau)}
$$
where, $C$ represents total codewords, $\mathbf{e}_c$ the codeword $c$ embedding, $\mathbf{W}$ a projection matrix, and $\phi(\tilde{\mathcal{X}})_t$ the output feature sequence at step $t$. The final prediction loss combines cross-entropy losses $L_m$ and $L_u$ over masked and unmasked timesteps, defined as:

$$
L_m(f ; \mathcal{X}, M, Z)=\sum_{t \in M} \log p_f\left(z_t \mid \tilde{\mathcal{X}}, t\right)
$$
where $L_u$ is similarly defined for $t \notin M$. To enhance representation learning, cluster ensembles provide supplementary information, and cluster assignments are refined by applying a new cluster generation trained over latent representations.

\subsection{Time-Frequency and Time Domain SE Front-ends}
\label{subsec:frontends}
\begin{figure*}[tp]
  \centering
  \includegraphics[width=\linewidth]{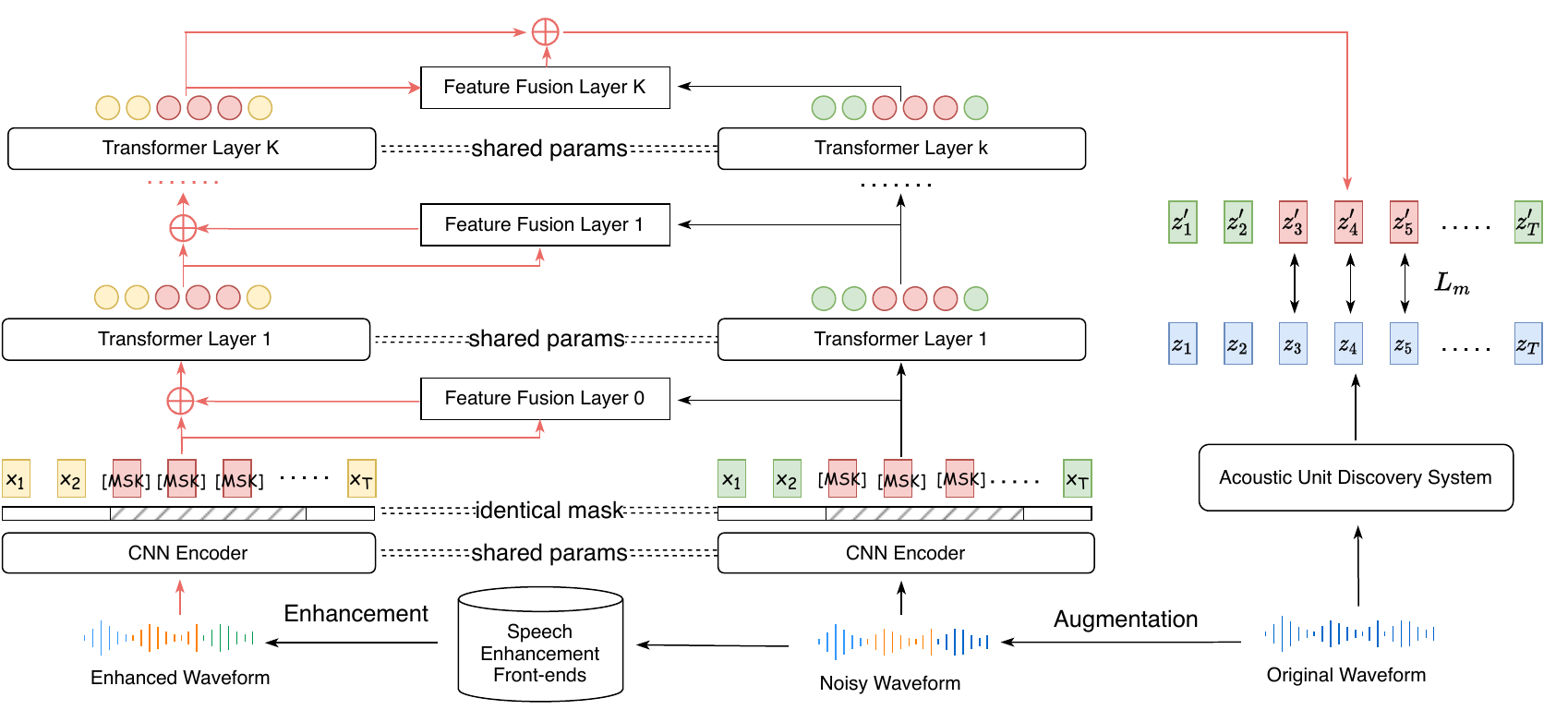}
  \caption{The proposed FAT-HuBERT framework: Data preprocessing via augmentation and enhancement with the IMST strategy, as detailed in Section \ref{subsec:intra_utterance_multi_style_training}. 
  Features derived from the enhanced waveform are integrated with those from the noisy waveform after the CNN encoder and each transformer layer. To protect the initialization parameters from premature continual pretraining, a residual connection is incorporated at each fusion stage. The HuBERT loss is computed on predicted codewords and the clustered codewords from original waveforms within the masked regions.}
  \vspace{-1em}
  \label{fig:fat}
\end{figure*}

Speech enhancement aims to enhance the quality of degraded speech signals. Mathematically, the observed signal $y(t)$ can be represented as the sum of a clean speech signal $x(t)$ and additional noise $n(t)$:
\begin{equation}
    y(t) = x(t) + n(t)
\end{equation}

The challenge lies in estimating the clean speech signal $\hat{x}(t)$ from the noisy signal $y(t)$. The effectiveness of the enhancement depends on the chosen representation for $y(t)$ and the specific approach used to generate $\hat{x}(t)$.

In the time-frequency~(TF) domain SE, the Discrete Fourier Transform (DFT) is employed to transform the time-domain signal into the frequency domain. This transformation provides both magnitude and phase components that can be utilized in the enhancement process. Techniques like the phase-sensitive mask~(PSM)~\cite{Hasannezhad2021PSM} have been developed to incorporate phase information. However, the transformation and reconstruction process can introduce errors, potentially impacting the quality of the resulting signal.

On the other hand, time-domain SE methods directly operate on the raw waveform of the noisy speech. For instance, adaptive front-end approaches~\cite{luo2019conv, luo2020dprnn} involve training an encoder module to transform the raw waveform into a latent representation. This representation is then processed by a separation module to extract individual signals, followed by reconstruction using a decoder module. Time-domain methods inherently incorporate phase information, avoiding transformation errors. However, they may have limitations in frequency representation~\cite{ochieng2022deep}, which can impact speech quality. Additionally, these methods often require more complex models due to the larger input space of raw waveforms.

\subsection{Front-end Adaptive Training~(FAT)}
\label{subsec:frontend_adaptive_training}
We propose the front-end adaptive training~(FAT) framework that utilizes a collection of pretrained speech enhancement models during the training phase to introduce diversity in training conditions and expose the system to a variety of distortions associated with different SE front-ends.

We incorporate models from both the time-domain and TF-domain. In the time-domain, we include solutions such as DPRNNTasnet~\cite{luo2020dprnn} and ConvTasnet~\cite{luo2019conv}, which are well known for their ability in signal reconstruction. In the TF-domain, we introduce the Deep Complex Convolutional Recurrent Network (DCCRNet)~\cite{hu2020dccrn}, Deep Complex Unet (DCUnet)~\cite{choi2018phase}, and Dual-path Transformer (DPTNet)~\cite{chen2020dpt}, recognized for their capability to perform advanced spectral transformations and reconstructions.

During training, our model adopts a dual inputs configuration as illustrated in Fig.\ref{fig:fat}. The first branch directly takes in the noisy waveform, while the second branch processes an enhanced waveform. The enhanced waveform is generated by applying a randomly selected front-end to the noisy waveform for each batch. Within the FAT framework, we introduce two techniques: intra-utterance multi-style training (IMST) and layer-wise feature fusion (LWFF).
% This approach leverages the diverse characteristics of different front-ends, thereby fostering an SSL model capable of effectively handling various distortion conditions.

\subsubsection{Intra-utterance Multi-style Training~(IMST)}
\label{subsec:intra_utterance_multi_style_training}
The intra-utterance multi-style training~(IMST) strategy is designed to build a more robust network by exploring various acoustic conditions within a single utterance.

During the training process, we consider a set of utterances $\{u_i\}_{i=1}^N$ per batch for the enhanced waveform branch, where $N$ is the batch size. For each utterance $u_i$, we select two distinct time intervals, denoted as $S_1 = \langle t_1, t_2 \rangle$ and $S_2 = \langle t_1^{\prime}, t_2^{\prime}\rangle$. Here, $S_1$ and $S_2$ represent segments of time in each utterance. Importantly, $S_1$ and $S_2$ are chosen identically for all utterances in a given batch. $S_1$ and $S_2$ can overlap, be disjoint, or contained within the other.

Instead of directly enhancing $u_i$, the utterance undergoes two steps: (1) \textbf{Augmentation}: The interval $S_1$ in $u_i$ is augmented with additive noise, creating a noisy segment $u_{i, S_1}^{\text{noisy}}$ within $u_i$. (2) \textbf{Enhancement}: The interval $S_2$ in $u_i$ is processed using a randomly selected SE front-end, yielding an enhanced segment $u_{i, S_2}^{\text{enhanced}}$ within $u_i$. The procedure is formally demonstrated in Algorithm~\ref{algo:fat}. Volume normalization is applied to the modified segments in the fifth and eighth lines.

IMST ensures that each utterance within a batch contains clean, noisy, and enhanced segments, thereby providing the network with a multi-style input for learning. The benefits of IMST are two-fold:
(1) Despite many SE front-ends being demanding in terms of GPU memory and computation time, IMST efficiently retains the training speed through the strategic selection of intervals for augmentation and enhancement. (2) IMST promotes the learning of a more robust representation by incorporating diverse acoustic conditions within the same utterance.

\begin{algorithm}
\DontPrintSemicolon
\KwIn{Training dataset $D = \{u_i\}_{i=1}^{N}$, Batch size $B$}
\KwIn{Noise dataset $\mathcal{N} = \{n_j\}_{j=1}^{M}$}
\KwIn{Set of SE front-ends $\mathcal{F} = \{f_k\}_{k=1}^{K}$}
Select $S_1 = \langle t_1, t_2 \rangle$ and $S_2 = \langle t_1^{\prime}, t_2^{\prime} \rangle$ for all $u_i \in b = \{u_{i}\}_{i=1}^{B}$\;
\For{$u_i \in b$} {
    $n \leftarrow \text{random\_sample}(\mathcal{N})$\;
    $u_{i,S_1}^{\text{noisy}} \leftarrow u_{i,S_1} + n$\;
    $u_{i,S_1}^{\text{noisy}} \leftarrow \frac{u_{i,S_1}^{\text{noisy}}}{\|u_{i,S_1}^{\text{noisy}}\|}\|u_{i,S_1}\|$\;
    $f \leftarrow \text{random\_sample}(\mathcal{F})$\;
    $u_{i,S_2}^{\text{enhanced}} \leftarrow f(u_{i,S_2})$\;
    $u_{i,S_2}^{\text{enhanced}} \leftarrow \frac{u_{i,S_2}^{\text{enhanced}}}{\|u_{i,S_2}^{\text{enhanced}}\|}\|u_{i,S_2}\|$\;
}
\caption{Data processing with IMST}
\label{algo:fat}
\end{algorithm}

\subsubsection{Layer-wise Feature Fusion~(LWFF)}
\label{subsec:layerwise_feature_fusion}
As illustrated in Fig.\ref{fig:fat}, a feature fusion module is employed to integrate features from the enhanced and noisy branches at each layer. We explore three distinct types of fusion modules:

\textbf{Observation Adding (OA):} Drawing inspiration from \cite{iwamoto2022bad}, where a scaled version of the observed signal is added to the enhanced speech to increase the signal-to-artifact ratio~(SAR), we apply OA within the latent space for features:
\begin{equation}
    Z_{\text{OA}} = Z_{\text{en}} + \alpha \cdot Z_{\text{noisy}},
\end{equation}
where $Z_{\text{OA}}$, $Z_{\text{en}}$, and $Z_{\text{noisy}}$ denote the fused, enhanced, and noisy features, respectively. $\alpha$ is a learnable scaling factor.

\textbf{Stacked Fusion (SF):} Features from the enhanced and noisy signals are stacked and mapped back to the original feature dimensions by a fully connected layer:
\begin{equation}
    Z_{\text{SF}} = \text{FC}( [ Z_{\text{en}} ; Z_{\text{noisy}} ] ),
\end{equation}
where $Z_{\text{SF}}$ is the fused feature map, FC represents the fully connected layer, and $[ ; ]$ signifies concatenation.

\textbf{Dual Attention (DA):} The DA module, proposed in \cite{zhu2022joint}, is applied in a layer-wise manner:
\begin{equation}
\begin{aligned}
Z_{\text {DA }}= & \text { Linear }\left(\operatorname{Multihead}\left(Z_{\text {en }}, Z_{\text {noisy }}, Z_{\text {noisy }}\right)\right)+ \\
& \text { Linear }\left(\operatorname{Multihead}\left(Z_{\text {noisy }}, Z_{\text {en }}, Z_{\text {en }}\right)\right)
\end{aligned}
\end{equation}
Here, $Z_{\text{DA}}$ denotes the fused feature map, while Multihead refers to the multi-head attention mechanism~\cite{vaswani2017attention}.

%%%%%%%%%%%%%%%%%%%%%%%%%%%%%%%%%%%%%%%%%%%%%
%%%%%%%%%%%%%%%% Experiments %%%%%%%%%%%%%%%%
%%%%%%%%%%%%%%%%%%%%%%%%%%%%%%%%%%%%%%%%%%%%%

\section{Experimental Setup}
\label{sec:experiments}
\subsection{Datasets}

We validate the effectiveness of the proposed method with both simulated and real-world noisy data. The preparation of these datasets employs three widely-used corpora in the field of ASR: LibriSpeech~\cite{librispeech}, WHAM!~\cite{wham}, and the 1-channel track of CHiME-4~\cite{chime4}. We denote the data employed for continual pretraining and fine-tuning of FAT-HuBERT as $D_P$ and $D_F$, respectively. The pretraining data, $D_P$, is synthesized by mixing the full 960 hours of LibriSpeech with noise randomly chosen from WHAM! at Signal-to-Noise Ratios (SNRs) uniformly sampled between 5 to 10 dB.

For testing on simulated noisy data, the original LibriSpeech \texttt{train-clean-100} partition is utilized for $D_F$. The original \texttt{test-clean} and \texttt{test-other} partitions are prepared for testing. Furthermore, by mixing the original test sets with noise from WHAM! at varying SNRs, we generate several simulated noisy test sets.

For testing on real-world noisy data, all data from CHiME-4, excluding the second channel due to its inferior quality, are used for fine-tuning. For testing, we resort to the official CHiME-4 1-channel real dev and eval sets.

\subsection{Speech Enhancement Front-ends}

In the pretraining phase, we incorporate five distinct front-ends: ConvTasnet, DCUNet, DPRNN Tasnet, DCCRN, and DPTNet, as described in Section \ref{subsec:frontend_adaptive_training}. These front-ends are trained on simulated data created by mixing noise from the WHAM! dataset with speech from \textsc{LibriSpeech} uniformly sampled betwwen 0 to 5 dB. All front-ends are trained with ESPnet ~\cite{watanabe2018espnet} default configs~\footnote{\url{https://github.com/espnet/espnet/tree/master/egs2/chime4/enh1/conf/tuning}}\footnote{\url{https://github.com/espnet/espnet/tree/master/egs2/wsj0\_2mix/enh1/conf/tuning}}.

During the testing phase, in addition to the five front-ends, we introduce two front-ends unseen during training: a SKiM~\cite{skim} front-end in the time-domain, and a 
 BLSTM~\cite{chen15o_interspeech} front-end in the TF-domain. When evaluating on the CHiME-4 dataset, all front-ends are trained using the simulated data from CHiME-4's 1-channel track.

\subsection{Pretraining}

We carry out pretraining with the \textsc{fairseq} toolkit~\cite{ott2019fairseq}. The adopted architectural design aligns with the one detailed in \cite{hubert}, which incorporates 12 transformer blocks, each having a hidden dimension of 768 and 8 heads. To expedite convergence, all models are initialzed with the  officially released HuBERT \textsc{Base} \footnote{\url{https://dl.fbaipublicfiles.com/hubert/hubert_base_ls960.pt}} checkpoint. Further, we employ k-means clustering with 500 clusters on the latent features extracted from the ninth layer of the official HuBERT \textsc{Base} model due to its superior phone purity as illustrated in \cite{hubert}. The continual pre-training shares the same configuration employed in training the second iteration of the HuBERT \textsc{Base} model, except that a lower learning rate of 1e-4 and fewer training steps 50k are applied unless indicated otherwise. 

\subsection{Model Fine-tuning}

Given that the CHiME-4 training set, which spans 92.28 hours, and the \textsc{LibriSpeech} 100-hour split are similar in terms of duration, we employ the \texttt{base\_100h} configuration from wav2vec 2.0 for both experiments.

\subsection{Decoding and Language Modeling}

For the \textsc{LibriSpeech} simulated and original test sets, we report the viterbi decoding result without an external language model. For the CHiME-4 test sets, a word-level language model based on LSTM is trained on the text part of the WSJ corpus~\cite{wsj} with the Espresso recipe~\cite{espresso}.

\section{Results and Analysis}
\label{sec:results}
\subsection{Results on Simulated Noisy Speech}

\begin{table*}
\caption{WER (\%) result on \textsc{LibriSpeech} (test clean / test other) simulated noisy datasets. The second line indicates the enhancement front-end applied during inference.}
\centering
\renewcommand{\arraystretch}{1.1}
\begin{adjustbox}{max width=\textwidth}
\begin{tabular}{l| c c c c c | c c c c c}
\hline \hline\noalign{\smallskip}
\multirow{2}{*}{Method} & \multicolumn{5}{c|}{0 $\sim$ 5dB test clean / test other} & \multicolumn{5}{c}{-5 $\sim$ 0dB test clean / test other} \\
& ConvTasNet & DCUNet & SkiM & BLSTM & NoEnh & ConvTasNet & DCUNet & SkiM & BLSTM & NoEnh \\
\noalign{\smallskip}\hline
1. Baseline &  19.1/37.1 & 16.4/33.3 & 20.8/41.5 & 27.0/49.0 & 38.7/60.2 & 34.6/57.8 & 31.9/53.1 & 34.5/64.5 & 53.4/71.8 & 70.3/84.2  \\
2. $+$ IMST &  16.7/32.9 & 15.2/29.8 & 17.1/33.3 & 21.6/38.6  & 18.2/34.4 & 29.0/48.1 & 26.8/43.8 & 29.6/48.3 & 41.5/57.6 & 34.4/51.7  \\
\noalign{\smallskip}\hline
3a. $++$ OA\_all &  14.0/28.9 & 12.8/27.4 & 15.1/29.2 & 16.4/32.6 & 17.1/34.0 & 26.1/44.4 & 23.9/41.4 & 27.6/44.7 & \textbf{34.7/51.5} & 35.6/52.3   \\
3b. $++$ OA\_first &  \textbf{13.2/28.9} & \textbf{12.4/26.6} & \textbf{13.9/29.2} & 17.8/35.0 & 16.6/33.9 & \textbf{24.2/43.7} & \textbf{22.8/40.5} & \textbf{25.0/43.9} & 38.0/54.0 & 35.5/53.1   \\
3c. $++$ OA\_last &  14.8/29.8 & 13.1/28.0 & 15.5/30.6 & 17.1/34.1 & 16.4/33.1 & 28.2/46.3 & 25.6/43.4 & 29.3/46.7 & 36.8/53.9 & 34.3/\textbf{50.9} \\
\noalign{\smallskip}\hline
4a. $++$ SF\_all &  40.5/57.8 & 40.6/57.7 & 40.8/57.6 & 40.9/57.3 & 40.5/57.6 & 66.0/76.0 & 65.8/75.4 & 65.8/75.8 & 66.1/75.9 & 66.1/75.9  \\
4b. $++$ SF\_first  & 14.0/29.3 & 12.6/26.7 & 15.0/29.7 & 18.0/34.6 & 18.7/35.5 & 25.5/43.7 & 23.3/40.5 & 26.3/44.1 & 37.8/54.0 & 37.9/54.4  \\
4c. $++$ SF\_last &  13.6/29.3 & 12.8/27.4 & 14.4/29.6 & \textbf{16.2/32.8} & \textbf{16.3/32.7} & 26.6/45.7 & 25.0/43.4 & 28.3/46.1 & 34.9/52.3 & \textbf{33.5}/51.0  \\
\noalign{\smallskip}\hline
5a. $++$ DA\_all &  46.4/64.1 & 46.3/64.0 & 46.3/64.1 & 46.4/64.3 & 46.4/64.0 & 69.8/80.1 & 69.6/80.2 & 69.5/80.1 & 69.8/80.0 & 69.9/80.1  \\
5b. $++$ DA\_first &  14.4/29.4 & 12.8/26.7 & 15.1/30.6 & 18.7/35.3 & 20.3/36.4 & 26.2/44.5 & 24.1/40.9 & 27.5/45.1 & 38.8/55.1 & 40.2/55.6  \\
5c. $++$ DA\_last &  14.8/30.6 & 18.5/28.5 & 15.9/31.3 & 17.5/34.5 & 17.3/33.8 & 29.6/47.7 & 27.9/45.6 & 31.4/48.2 & 38.8/55.0 & 36.9/53.0  \\
\bottomrule
\end{tabular}
\end{adjustbox}
\label{tab:ls_simu}
\end{table*}

Table~\ref{tab:ls_simu} presents the performance of the fine-tuned SSL model on simulated test sets derived from the \textsc{LibriSpeech} dataset, enhanced by various SE front-ends. The table includes SE front-ends used for pretraining, as well as front-ends unseen during pretraining, encompassing both time-domain and TF-domain models. The baseline model is a HuBERT model finetuned on the \textsc{LibriSpeech} \texttt{ls-clean-100} partition. Since the FAT-HuBERT model takes two branches as input, both branches are fed with the same noisy waveform in the NoEnh columns.

Comparing the rows denoted as *.a with those denoted as *.b and *.c, applying fusion on all layers generally leads to inferior performance compared to restricting fusion to either the first or final layer. 
This is especially noticeable with the DA module, possibly due to the extra parameters it introduces. Despite residual connections are applied, these extra parameters still have an impact on the initial parameters of the pretrained HuBERT model during continual pretraining.

Fusion at either the initial or final layer consistently outperforms the IMST approach, highlighting the effectiveness of the fusion module in mitigating distortions introduced by the SE front-ends. Notably, the first-layer for OA fusion and the final-layer for DA and SF fusion yield the best results. This distinction can be attributed to the minimal parameter introduction of OA ($\alpha$), which has a smaller impact on the performance of upper layers during pretraining.

The benefits of the fusion module extend beyond the training front-ends, as it demonstrates generalization capability to unseen front-ends. Furthermore, the fusion module exhibits robust performance even under lower SNRs than those encountered during front-end training, indicating its ability to handle challenging conditions effectively.

\begin{figure}
  \centering
  \includegraphics[width=\linewidth]{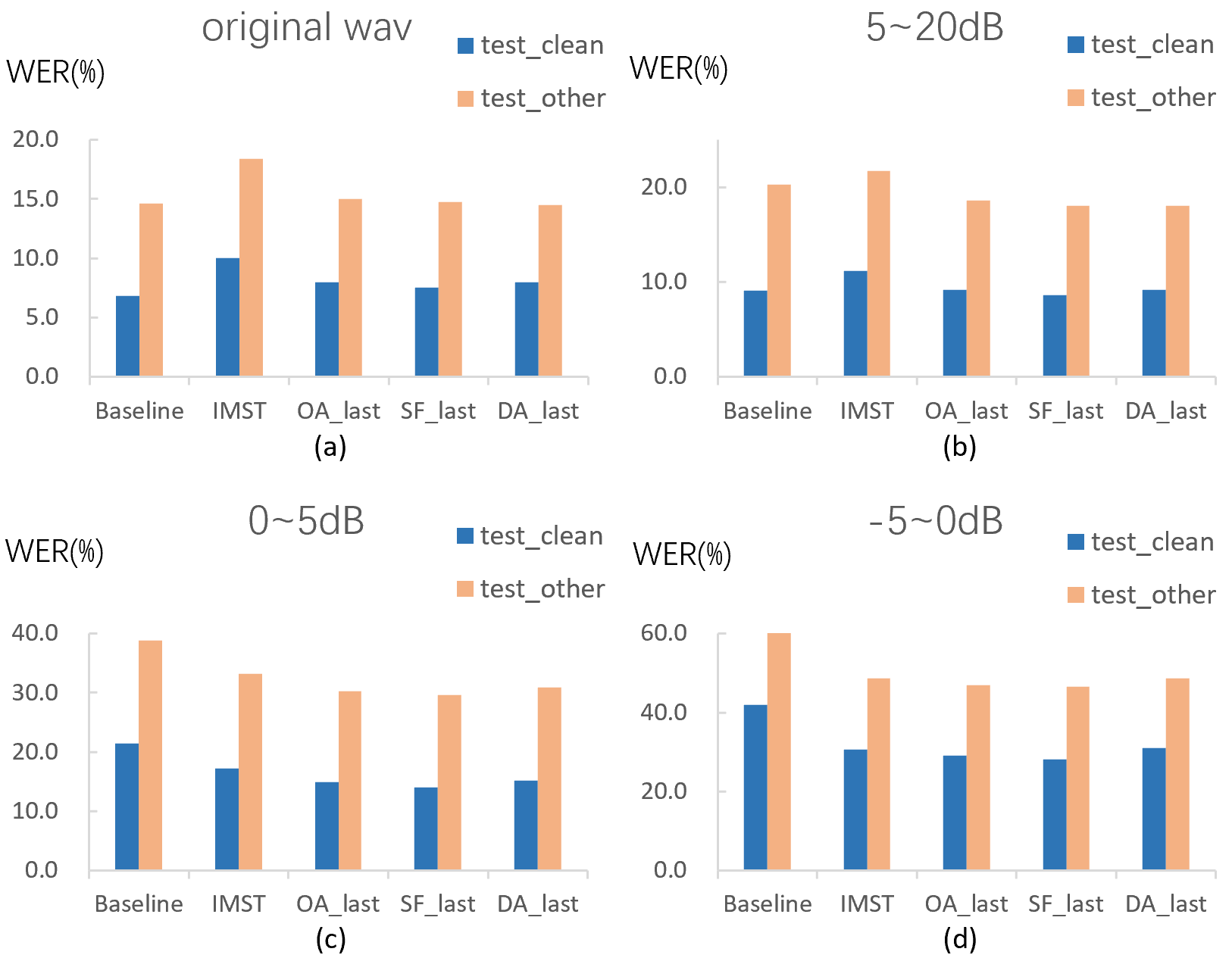}
  \caption{WER(\%) on original test data (test clean / other) and simulated noisy data under different SNR conditions.}
  \label{fig:wer}
\end{figure}

\subsection{Performance at different SNR conditions}

Figure \ref{fig:wer} presents the WER results of the proposed method under different SNR conditions. The results are obtained by averaging the WER across all seven different front-ends used for enhancement. Notably, in Figure \ref{fig:wer} (a), the original waveform is enhanced without being mixed with noise, demonstrating the robustness of the fusion module  under clean conditions.

As depicted in Figures \ref{fig:wer} (c) and (d), the IMST strategy effectively enhances the model's performance in low SNR conditions. However, under high SNR conditions illustrated in Figures \ref{fig:wer} (a) and (b), a slight degradation in performance is observed when IMST is applied, suggesting its limited utility in less noisy conditions.

On the other hand, the fusion module is beneficial in mitigating the performance degradation associated with IMST under high SNR conditions. Furthermore, it brings about additional WER reduction in low SNR conditions.

\subsection{Results on Real-World Noisy Speech}

In this section, our proposed methods are evaluated on the CHiME-4 1-channel real noisy test sets as shown in Table~\ref{tab:chime4}. We include recently published SSL results~\cite{wav2vec-switch, wav2vec-recons,wang2023hubert}, which are also SSL based robust ASR models.

\begin{table}[htbp]
\centering
\renewcommand{\arraystretch}{.9}
%\centering
\caption{WER(\%) of different systems on CHiME-4 1-channel real test sets }
\begin{tabular}{l|c  | c  c }
\hline\noalign{\smallskip}
\multirow{2}{*}{System} & \multirow{2}{*}{front-end} & \multicolumn{2}{c}{CHiME-4 REAL} \\
&  & dev & eval \\
\midrule
Yang et al. \cite{Yang2022ACB} & \multirow{3}{*}{N/A} & 3.4  & 6.3  \\
wav2vec-switch \cite{wav2vec-switch} &  & 3.5  & 6.6  \\
wav2vec (recons) \cite{wav2vec-recons} &  & 5.0  & 9.0  \\
\midrule
\multirow{5}{*}{\makecell{HuBERT-AGG~\cite{wang2023hubert} \\ (50k steps)}} & N/A & 3.3 & 6.1 \\
& ConvTasNet & 3.4 & 6.2 \\
& DCUNet & 3.5 & 6.4 \\
& SKiM & 3.3 & 6.0 \\
& BLSTM & 3.7 & 6.8 \\
\midrule
\multirow{5}{*}{HuBERT} & N/A & 4.4 & 8.6 \\
 & ConvTasNet & 4.6 & 8.7 \\
 & DCUNet & 4.1 & 8.3 \\
 & SKiM & 3.9 & 7.7 \\
 & BLSTM & 4.3 & 8.4 \\
\midrule
\multirow{4}{*}{$+$IMST} & ConvTasNet  & 4.1 & 8.2 \\
 & DCUNet  & 4.0 & 7.7 \\
 & SKiM  & 3.8 & 7.4 \\
 & BLSTM  & 4.0 & 8.1 \\
\midrule
$++$ OA\_first & \multirow{3}{*}{SKiM} & 3.3 & 5.9 \\
$++$ SF\_first &  & \textbf{3.1} & \textbf{5.7} \\
$++$ DA\_first &  & 3.5 & 6.3 \\
\bottomrule
\end{tabular}
% \end{adjustbox}
\label{tab:chime4}
\end{table}

Results on the CHiME-4 1-channel real-word test sets are presented in Table~\ref{tab:chime4}. Results for HuBERT and HuBERT-AGG indicates directly prepending a SE front-end does not necessarily improves ASR performance. The application of IMST exhibits a consistent enhancement over the baseline. It is worth noting that this improvement is achieved without joint tuning of the SSL backend and the independently retrained front-ends using CHiME-4 data. Moreover, the integration of the fusion module contributes additional enhancements, with the SF fusion module demonstrating notable benefits.

%%%%%%%%%%%%%%%%%%%%%%%%%%%%%%%%%%%%%%%%%%%%%
%%%%%%%%%%%%%%%% Conclusions %%%%%%%%%%%%%%%%
%%%%%%%%%%%%%%%%%%%%%%%%%%%%%%%%%%%%%%%%%%%%%

\section{Conclusions}
\label{sec:conclusions}
% In this study, we proposed the Front-End Adaptive Training (FAT) approach, which takes advantage of a variety of pretrained speech enhancement models to diversify the training acoustic conditions.
% We propose Intra-Utterance Multi-Style Training (IMST) strategy, which ensures exposure to a variety of acoustic conditions within an individual utterance. This technique proved valuable in scenarios with low SNRs, although its effectiveness was somewhat curtailed under high SNR circumstances.
% To address this limitation, we developed the Layer-Wise Feature Fusion (LWFF) technique, a dedicated method that merges features from both the enhanced and noisy branches. This technique efficiently counteracted performance deterioration in high SNR scenarios, and consistently improved results across a variety of front-ends.
% We conduct experiments using \textsc{LibriSpeech} simulated noisy speech and \textsc{CHiME-4} real-world noisy speech, demonstrating significant reductions in WER and validating the overall effectiveness and generalizability of our methods. 

In this study, we introduced the Front-End Adaptive Training (FAT) approach, utilizing a multitude of diverse pretrained speech enhancement models to adapt SSL model to various kinds of distortions introduced by SE front-ends. We propose Intra-Utterance Multi-Style Training (IMST) strategy, which proved effective in low SNR scenarios but exhibited limitations under high SNR circumstances. To address this, we present the Layer-Wise Feature Fusion (LWFF) method, mitigating performance deterioration in high SNR scenarios and consistently improving results across different front-ends. Our experiments on simulated and real-world noisy speech from \textsc{LibriSpeech} and \textsc{CHiME-4} respectively demonstrated significant performance improvement, validating the effectiveness of our methods.

\section{acknowledgement}
This work was supported in part by China NSFC projects under Grants 62122050 and 62071288, and in part by Shanghai Municipal Science and Technology Major Project under Grant 2021SHZDZX0102.

\bibliographystyle{IEEEbib}
\bibliography{refs}

\end{document}